\documentclass[12pt]{article}
\usepackage{amsmath}
\usepackage{amssymb}
\usepackage[titletoc,title]{appendix}
\numberwithin{equation}{section}
\usepackage[left=0.98in,right=0.98in,bottom=1.5in]{geometry}
\usepackage{setspace}
\usepackage [linktocpage]{hyperref}
\begin{document}
\title{\bf Non-Minimally Coupled Pseudoscalars \\ in AdS$_4$ for Instantons in CFT$_3$ \ \\ \
\\ }
\author{{\bf M. Naghdi \footnote{E-Mail: m.naghdi@mail.ilam.ac.ir} } \\
\textit{Department of Physics, Faculty of Basic Sciences}, \\
\textit{University of Ilam, Ilam, West of Iran.}}
\date{\today}
 \setlength{\topmargin}{0.0in}
 \setlength{\textheight}{9.2in}
  \maketitle
  \vspace{0.0in}
    \thispagestyle{empty}
 \begin{center}
   \textbf{Abstract}
 \end{center}
For the 11-dimensional supergravity over $AdS_4 \times S^7/Z_k$, beginning with a general 4-from ansatz and the main geometry
unchanged,
we get a tower of massive and tachyonic pseudoscalars. Indeed, the resultant equations can be assigned to the so-called
$\phi^4$ actions
of the non-minimally coupled scalar-tensor theories with a cosmological constant. We focus on a well-known tachyonic and a
new
massive
bulk mode, which are singlet under the internal group and break all supersymmetries, associated with skew-whiffing and
Wick-rotating of the
background 4-from flux, respectively. The first one is the conformally coupled $m^2=-2$ pseudoscalar in the bulk of Euclidean
$AdS_4$,
where an exact instanton solution is found and a marginally triple-trace deformation with a proper dimension-1 operator
produces an
agreeing boundary solution with finite action. From the action evaluated on the solution, we estimate the decay rate of the
vacuum
tunneling mediated by the instanton. Another massive $m^2=+4$ mode, with the so-called non-minimal coupling parameter $\xi =
-
1/3$, also
breaks the conformal invariance and so, there is no exact solution. Then, based on the AdS$_4$/CFT$_3$ correspondence rules,
we propose
the dimension-4 ($\Delta_+ = +4$) boundary operator in the skew-whiffed (anti-M2-branes) theory to deform the boundary
action-
consisting
of a singlet fermion, an original scalar and $U(1)$ gauges fields- with and find some solutions to be matched with the bulk
solutions.

\newpage
\setlength{\topmargin}{-0.7in}
\pagenumbering{arabic} 
\setcounter{page}{2} 

\section{Introduction}
Instantons as the fully localized objects (or point particles) in space can mediate various vacuum tunneling with important
roles in
physics of various field and gravity to early universe (inflationary) theories. Especially, in the gauge/gravity dualities,
they have
been used to perform nonperturbative tests and learn the facts from one side for another side of the duality; Look at
\cite{VandorenNieuwenhuizen} and references therein. After presenting a standard M2/D2-branes model by
Aharony-Bergman-Jafferis-Maldacena (ABJM) \cite{ABJM}, we have found some instantons and other localized objects for the
AdS$_4$/CFT$_3$ correspondence in \cite{I.N}, \cite{N}, \cite{Me3}, \cite{Me4} and \cite{Me5}.

The ABJM Lagrangian describes the world-volume action of N intersecting M2-branes on a $Z_k$ orbifold of $C^4$, where the
orbifold acts
as $y_A \rightarrow e^{i \frac{2 \pi}{k}} y_A$ on four complex coordinates $y_A$ with $A=1,2,3,4$. In the large N limit and
fixed $N/k$,
the 11-dimensional (11d) supergravity over $AdS_4 \times S^7/Z_k$ is reliable when $N\gg k^5$. This $\mathcal{N}=6$ conformal
$U(N)_k
\times U(N)_{-k}$ Chern-Simons-matter theory includes the gauge fields $A_i$ and $\hat{A}_i$, and the bifundamental scalars
$Y^A$ and
fermions $\psi_A$ that transforms as $\textbf{4}_1$ and $\bar{\textbf{4}}_{-1}$ under the subgroup $SU(4)_R \times U(1)_b$ of
the
original $SO(8)$ group. The latter isometry is valid for $k=1,2$, where the supersymmetry is also enhanced to $\mathcal{N}=8$
because of
monopole operators \cite{ABJM}.

To study and find instantons in the model, the main tool is the state-operator correspondence as settled, for instance, in
\cite{KlebanovWitten}. Here, we introduce a general 4-from ansatz for the 11d supergravity and then from the resulting
equations, deduce
some massless, massive and tachyonic scalars and pseudoscalars. Among them we concentrate on two singlet modes in the bulk of
Euclidean
$AdS_4$ ($EAdS_4$), which are supposed to come from wrapping the included (anti)M-branes around some internal $S^7/Z_k$
directions. The
first one is the well-known $m^2=-2$ conformally coupled (c.c.) pseudoscalar and the second one is a non-minimally coupled
(n.m.c.)
pseudoscalar with $m^2=+4$; and the so-called coupling to gravity are $\xi=1/6$ and $\xi=-1/3$, respectively. For the c.c.
case we write
an exact solution already studied also in \cite{I.N} while for the n.m.c. case we describe approximate methods and solutions.

Indeed, for the (pseudo)scalars with the masses around the so-called Breitenlohner-Freedman (BF) bound
\cite{BreitenlohnerFreedman} $m^2
\geq - \frac{9}{4}$ in the bulk of $AdS_4$, one may consider various boundary conditions with preserving the asymptotic AdS
symmetries (the
so-called designer gravity theories). As a result, the bulk solutions with a big crunch singularity and AdS black holes with
scalar hairs are also founded; Look for instance at \cite{HertogMaeda01}. Especially, we see that with a triple-trace
deformation for the
c.c. case, the instanton solution causes instability and tunneling among the vacua of the model with breaking all
supersymmetries. In
addition, for the n.m.c. case, with breaking the conformal invariance, there is not any known regular and exact Euclidean
bulk
solution
with finite action. Meanwhile, with broken conformal symmetry and so $EAdS_4$ isometry from $SO(4,1)$ to $SO(4)$,
which the latter is the isometry of $S^3$ used also for the boundary space here, there is an infinite family of boundary
instantons
causing instability. With these setups, one can also achieve the holographic descriptions for cosmological singularities;
Look
for
instance at \cite{SmolkinTurok}.

Anyhow, the solutions here are $SU(4) \times U(1)$-singlet and break all 32 supersymmetries. Therefore, to do the right
bulk-boundary
correspondence, we should reshuffle the original representations $\textbf{8}_s$, $\textbf{8}_c$ and $\textbf{8}_v$ for
supercharges,
fermions and scalars of the main M2-branes theory, respectively; Look at \cite{NilssonPope} and \cite{Duff}. Indeed, the
skew-whiffing
$\textbf{8}_s \leftrightarrow \textbf{8}_c$ meets our purpose and so, the resultant theories will be for anti-M2-branes. One
the
boundary 3d field theory, we deform the actions with some suiting dimension-1 and -4 operators besides mixed and Dirichlet
boundary terms
for the c.c. and n.m.c. case respectively, and get the solutions with finite actions with adjustments of AdS$_4$/CFT$_3$
duality.

The rest of this note is organized as follows. In Section 2, we go with the gravity side of the study. There, we discuss the
background,
ansatz and aspects of the solutions and, in subsection 2.4, the correction from the instanton solution for the c.c.
pseudoscalar is
computed. Meanwhile, we hint on the supersymmetry breaking for the latter case in Appendix A, briefly. Section 3 is devoted
to
the field
theory side of the study, where the dual boundary counterparts for both bulk modes, based on the symmetries and other
gauge/gravity
correspondence rules, are established. After we discussed the basic bulk-boundary correspondence in subsection 3.1, in
subsection 3.2,
the dual boundary instantons for the c.c. bulk pseudoscalar will be built with a triple-trace deformation and a proposed
dimension-1
operator. There will also be discussions on the boundary effective action and decay rate of the unstable vacuum. In
subsection
3.3, we
set up the agreeing dimension-4 boundary operator for the n.m.c. bulk pseudoscalar; and then deform the action with suitable
boundary
terms to arrive at the solutions with finite actions to meet the bulk constrained instantons. In Section 4, we present a
summary with
comments on the issues to be addressed further.

\section{The 11-Dimensional Gravity Aspects}

\subsection{Some Preliminaries}
We use the supergravity metric
\begin{equation}\label{eq001}
ds^2_{11d}=\frac{R^2}{4} ds^2_{EAdS_4} + R^2 ds^2_{S^7/Z_k},
\end{equation}
with $R=R_7=2R_{AdS}=2L$ for the 11d tangent-space radius of curvature and
\begin{equation}\label{eq001a}
\ \ \ \ \ \ ds^2_{EAdS_4}=\frac{1}{u^2} \big(du^2+ dx_i dx_i \big), \quad i=1,2,3,
\end{equation}
for the Euclidean $AdS_4$ metric in upper-half Poincar$\acute{e}$ coordinate, and
\begin{equation} \label{eq001b}
ds_{S^7/Z_k}^2 =ds_{CP^3}^2+e_7\otimes e_7, \quad e_7=\frac{1}{k}(d\varphi+k\omega),
\end{equation}
where $S^7/Z_k$ is considered as a $U(1)$ fiber-bundle on $CP^3$ with the coordinate $\acute{\varphi}=\varphi/k$, and
$J(=d\omega)$ for
the K$\ddot{a}$hler form, with the topologically nontrivial 1-from $\omega$, on $CP^3$. \\
The background 4-form of ABJM \cite{ABJM} reads
\begin{equation}\label{eq002}
  G_4^{(0)}=d\mathcal{A}_3^{(0)} = \frac{3}{8} R^3 \mathcal{E}_4 =  N \mathcal{E}_4,
\end{equation}
with $\mathcal{E}_4$ as the unit-volume form of $AdS_4$ and $N$ units of the 4-flux on the quotient space.

From the Euclideanized 11d supergravity action \cite{Me5}, the equations of motion read
\begin{equation}\label{eq003}
  d \ast_{11} G_4 - \frac{i}{2} G_4 \wedge G_4=0,
\end{equation}
with $\ast_{11}\equiv \ast$ for the 11d Hodge-star and
\begin{equation}\label{eq004}
    \mathcal{R}_{MN} - \frac{1}{2} g_{MN} \mathcal{R}=\kappa_{11}^2 T_{MN}^{G_4},
\end{equation}
where $M, N,..$ are the full space-time indices, $\mathcal{R}$ is the scalar curvature, $\kappa_D^2=8 \pi \mathcal{G}_{D}$
with
$\mathcal{G}_{11}$ as the 11d Newton's constant, and $T_{MN}^{G_4}$ is the energy-momentum tensor of the 4-form flux.

\subsection{General 4-Form Ansatz and Equations}
We consider the combined ansatz, associated with some included (anti)M-branes, as
\begin{equation}\label{eq01}
G_4 = f_1\ G_4^{(0)} + df_2 \wedge \mathcal{A}_3^{(0)}+ \ast_4 df_3 \wedge d\varphi + f_4 \mathcal{A}_3^{(0)} \wedge d\varphi
+  df_5
\wedge e_7 \wedge J + f_6\ J^2,
\end{equation}
where $f_1, f_2,...$ are (pseudo)scalar functions in $EAdS_4$ space. From the Bianchi identity $dG_4=0$, we simply obtain
\begin{equation}\label{eq02a}
      f_5 = f_6,
\end{equation}
\begin{equation}\label{eq02b}
      d(\ast_4\ df_3) = 0 \Rightarrow  f_3(u,\vec{u}) = c_1 + \frac{c_2 u^3}{\left(u^2 + (\vec{u}-\vec{u}_0)^2\right)^3},
\end{equation}
\begin{equation}\label{eq02c}
      df_4 \wedge \mathcal{A}_3^{(0} + f_4\  G_4^{(0)}=0 \Rightarrow f_4(u)=c_3 {u^3},
\end{equation}
with $c_1, c_2,...$ as some bulk constants and $r=|\vec{u}|=\sqrt{x_i x^i}$. Next, to satisfy (\ref{eq003}) with $G_4$,
besides the
latter three conditions, we must first set
\begin{equation}\label{eq02d}
      f_2 = f_2(x,y,z);
\end{equation}
Then, the remaining relations to be satisfied are
\begin{equation}\label{eq03a}
     -i f_6 df_6 + c_4 df_1 \wedge \ast_4 G_4^{(0)}=0,
\end{equation}
\begin{equation}\label{eq03b}
     -i f_6 f_1 G_4^{(0)} + c_5 c_6 f_6 \mathcal{E}_4 - c_{5}^{-1} d(\ast_4\ df_6)=0,
\end{equation}
where, and in future we use,
\begin{equation}\label{eq03c}
      \ast_7 \textbf{1} = c_4 J^3 \wedge e_7, \quad \ast_7 (J^2) = c_5 J \wedge e_7,  \quad \ast_4
     \textbf{1} = c_6 \mathcal{E}_4, \quad \ast_7 e_7=\mathcal{E}_6= c_{7} J^3, 
\end{equation}
and note that the minus sign in the last term on LHS of (\ref{eq03b}) is due to $\varepsilon_{\mu m n p \nu \rho \sigma q r s
7}=
-\varepsilon_{\mu \nu \rho \sigma} \varepsilon_{m n p q r s 7}$ when doing the 11d star operation, with  $\mu, \nu, ... $ and
$m, n, ...
$ for the external and internal indices, respectively.

We notice that the solutions (\ref{eq02b}) and (\ref{eq02c}) are those already studied in \cite{Me3}, \cite{Me4} and
\cite{Me5}, while a
solution like (\ref{eq02d}) was introduced in \cite{N} and we focus on it more in future studies. However, one may also note
that the
last two equations (\ref{eq03a}), (\ref{eq03b}) are when we consider just the first, fifth and sixth terms of  the ansatz
(\ref{eq01}). Now, by changing
$f_1 N =\bar{f_1}$ for convenience, (\ref{eq03a}) reads
\begin{equation}\label{eq04}
    \bar{f_1}=i \frac{3}{8 R^3} \frac{f_6^2}{2} \left(\frac{R^{2d_2}}{R^{d_1}} \right) \pm i \bar{c}_1,
\end{equation}
in which $\bar{c}_1$ is some convenient constant and we have used the dimensional coefficients $R^{d_1}$ and $R^{d_2}$ for
the
first and
fifth-sixth terms of the ansatz respectively, and that
\begin{equation}\label{eq05a}
     c_4 = \frac{1}{3!} R^7, \quad c_5 = \frac{2}{R}, \quad  c_6 = \frac{R^4}{16}, \quad c_{7} = \frac{R^5}{3!}.
\end{equation}
Then, with $d_1=0$ and $d_2=4$ and $\bar{c}_1 = \bar{C} R^3$, with $\bar{C}$ as a rational number, one can obtain a tower (an
infinite
set) of massive and tachyonic bulk (pseudo)scalar modes (some deformations of the gravity background 4-form corresponding to
some deformations
of the boundary field theory). The interesting case is when $\bar{C}=\frac{3}{8}$ for which, with $f_6 \equiv f$ from now on,
we obtain
\begin{equation}\label{eq06}
\frac{1}{\sqrt{g_4}} \partial_{{\mu}} \left(\sqrt{g_4}\ g^{{\mu}{\nu}} \partial_{{\nu}} f \right)- \frac{4}{R^2} f \mp
\frac{12}{R^2} f
- 2
\times 3 f^3=0,
\end{equation}
where the lower ($+$) (the upper $-$) sign is for an exact skew-whiffing (Wick rotating) of the background (\ref{eq002}) and
corresponds to a conformally (non-minimally) coupled pseudoscalar $m^2 R_{AdS}^2=-2$ ($m^2 R_{AdS}^2=+4$) in the bulk of $EAdS_4$. So, the dual boundary operator corresponding to the normalizable bulk mode has the scaling dimension $\Delta_+=2$ ($\Delta_+=4$); we return to this issue soon.

Besides, it is notable that the last setup could be in general considered as a consistent reduction of the 11d supergravity to four dimensions \cite{deHaro2}, where the resultant 4d (the so-called $\phi^4$) action reads
\begin{equation}\label{eq07}
     S_4^E = \int{ d^4x\ \sqrt{g_4}\ \left(-\frac{1}{2{\kappa}_4^2}\left(\mathcal{R}_4 - 2\Lambda \right)+\frac{1}{2}g^{\mu
     \nu}
     ({\partial}_{\mu} f)({\partial}_{\nu}f) + \frac{1}{2} \xi \mathcal{R}_4 f^2 + V(f)\right)},
\end{equation}
in which
\begin{equation}\label{eq07b}
    V(f) = \frac{\lambda}{2} f^4, \quad \Lambda = - \frac{12}{R^2},
\end{equation}
where $\lambda$ is an arbitrary dimensionless coupling-constant, $\xi$ is the non-minimal coupling parameter, $\Lambda$ is
the cosmological constant and the index 4 on $g_4, {\kappa}_4$ and $\mathcal{R}_4$ is for $EAdS_4$. Further, from varying the action (\ref{eq07}) with respect to the metric $g_{\mu \nu}$, we obtain
\begin{equation}\label{eq09}
    {G}_{\mu \nu} + \Lambda g_{\mu \nu} = \kappa_4^2\ \mathcal{T}_{\mu \nu},
\end{equation}
in which ${G}_{\mu \nu}$ is the Einstein tensor and
\begin{equation}\label{eq09a}
  \begin{split}
   & \ \ \ \ \ \ \ \ \ \ \ \ \ \ \ \ \ \mathcal{T}_{\mu \nu} = T_{\mu \nu}^{m.c.} + T_{\mu \nu}^{imp.}, \\
   & T^{m.c.}_{\mu \nu} = {\nabla}_{\mu}f\ {\nabla}_{\nu}f - g_{\mu \nu}\big(\frac{1}{2} g^{\rho \sigma} {\nabla}_{\rho}f\
   {\nabla}_{\sigma}f
   + V(f)\big), \\
   & T^{imp.}_{\mu \nu} = \xi \left(g_{\mu \nu} \square_{g_4} - {\nabla}_{\mu} {\nabla}_{\nu} + {G}_{\mu \nu} \right) f^2,
  \end{split}
\end{equation}
where the energy-momentum tensor is divided into a minimally coupled ($\xi=0$) and an improved part because of the
non-minimal coupling ($\xi\neq 0$).\\
Now, because the energy-momentum tensor $\mathcal{T}_{\mu \nu}$ is clearly traceless, with respect to the scalar equation from the action, the equation (\ref{eq09}) implies that all possible solutions of the theory have the constant scalar curvature
\begin{equation}\label{eq09b}
    \mathcal{R}_4 = -\frac{48}{R^2},
\end{equation}
which is indeed the Ricci scalar of $EAdS_4$ space with $R_{AdS}$ radius (see (\ref{eq001})) and also for the vacuum solution of the theory because of $\Lambda$ in (\ref{eq07b}). One may now note that for $\xi=\frac{1}{6}$, the \emph{conformally coupled} case and for $\xi=-\frac{1}{3}$, the \emph{non-minimally coupled} case are achieved in (\ref{eq06}), with $\lambda=3$ for both. \footnote{In fact, the conformal holography comes from the fact that the $EAdS_4$ metric from (\ref{eq001a}) is related to the flat $R^4$ one as $g_{\mu \nu}=\Omega(u)^{-2} \eta_{\mu \nu}$ with $\Omega(u)=\frac{2 u}{R}$ and that $f=f^{c.c.}=\Omega(u) f^{m.c.}$, where $c.c.$ and $m.c.$ indicate the conformally and minimally (weakly) coupled (pseudo)scalars, respectively. This fact is also clear from the fourth term of the action (\ref{eq07}) with factor $\frac{1}{R^2}$ in front of the (pseudo)scalar quadratic function.}

\subsection{Solution Aspects: Bulk Instantons} \label{sub2.2}
For the energy-momentum tensor of the conformally coupled pseudoscalar, one can write \cite{Papadimitriou02}
\begin{equation}\label{eq09b}
    \mathcal{T}_{\mu \nu} = \frac{1}{3} \frac{f^3}{\left(1 - \xi{\kappa}_4^2 f^2 \right)} \left({\nabla}_{\mu} {\nabla}_{\nu}
    -
    \frac{1}{4}
    g_{\mu \nu} \square_{g_4} \right) f^{-1},
\end{equation}
which is useful for finding a suitable solution. In fact, we note to the so-called \emph{stealth configurations} as the
nontrivial
solutions with vanishing stress tensor \cite{Ayon-Beato}. The latter has in common with our solution a vanished stress tensor
likewise
\emph{'t Hooft instantons} with zero energy-momentum tensors. \footnote{Indeed, it can be checked that the correction through
$T_{\mu \nu}^{\tilde{G}_4}$ of (\ref{eq004}) with our ansatz
\begin{equation}\label{eq01b}
\bar{G}_4= \bar{f}_1 \mathcal{E}_4 + R^4 df \wedge e_7 \wedge J + R^4 f\ J^2,
\end{equation}
with respect to the equation (\ref{eq06}) vanishes while the internal components do not and so, one cannot uplift the 4d
solution
to the full 11d one. Although one should include the backreaction in general analyses, for the purposes in studying the near
boundary
behaviors with probe approximations, we simply ignore the backreactions on the main geometry \cite{Skenderis}.}

Now, from vanishing (the ten equations) of the modified energy-momentum tensor (\ref{eq09b}), with covariant derivatives and
Chirstoffel symbols for $EAdS_4$ in the case, along with the pseudoscalar equation (\ref{eq06}), one can get a solution
directly. Alternatively, we use the conformal flatness of the external metric and then, from (\ref{eq09b}), the solution
reads
\begin{equation}\label{eq10}
f^{m.c.}(u,\vec{u}) = \frac{\tilde{b}_0} {\left[-b_0^2 + (a_0+u)^2 + (\vec{u}-\vec{u}_0)^2 \right]}, \quad
b_0^2=\frac{\lambda}{4}\tilde{b}_0^2,
\end{equation}
where the constraint on the RHS comes from the equation $\square_4 f^{m.c.} - 2\lambda (f^{m.c.})^3 = 0$, which is in turn
arisen from (\ref{eq06}) for the conformally coupled case, with $\square_4$ for Laplacian of the flat 4d Euclidean space.
\footnote{It is also mentionable that the solution for the massless case of this $\phi^4$ model is the so-called Fubini
instanton \cite{Fubini1}, which describes the tunneling without barrier from top of the potential to any arbitrary state; See
also \cite{Castel1}, \cite{TagirovTodorov} and \cite{Loran1}.} Then, from conformal property of $f$ (by multiplying $2u/R$ in
(\ref{eq10})), we obtain the exact solution
\begin{equation}\label{eq10a}
f(u,\vec{u}) = \frac{4}{R \sqrt{\lambda}} \left( \frac{b_0 u}{-b_0^2 + (a_0+u)^2 + (\vec{u}-\vec{u}_0)^2} \right),
\end{equation}
in which $a_0, b_0, u_0^i$ are some arbitrary constants and $a_0^2 > b_0^2 > 0$ to have a nonsingular solution; Look also at
\cite{I.N} for a similar derivation. This is a nontrivial solution (of instanton type) of the coupled equations (\ref{eq06})
and (\ref{eq09b}) with vanishing energy-momentum tensor. In fact, as we will see, $\vec{u}_0$ and $a_0^2-b_0^2$ parameterize
the 3d instanton vacuum with $b_0$ for the instanton's size and $\vec{u}_0$ for its location on the boundary.

On the other hand, for the non-minimally coupled case, because of the conformal symmetry breaking, one might not be able to
find an exact
solution. But, one can solve the corresponding equation
\begin{equation}\label{eq11}
\partial_{u}\partial_{u} \tilde{f} + \delta_{ij} \partial^{i}\partial^{j} \tilde{f} -\frac{6}{u^2} \tilde{f} -2 \lambda
\tilde{f}^3=0,
\quad
f = \frac{2 u}{R} \tilde{f},
\end{equation}
perturbatively to get an approximate solution with specials mathematical methods, which can in turn be matched with
\emph{constrained
instantons} in language of \cite{Affleck1}. Indeed, we can write, for (\ref{eq06}) with the upper sign, an iterative solution
like
\cite{HokerFreedman1}
\begin{equation}\label{eq12}
f(u, \vec{u})=f_0(u, \vec{u})- 2 \lambda \int{dw\ d^3\vec{w} \sqrt{g_4}\ G(u,\vec{u}; w,\vec{w}) f_0(w, \vec{w})^3},
\end{equation}
with
\begin{equation}\label{eq12a}
f_0(u, \vec{u})= \int d^3\vec{u}_0\ K_4(u,\vec{u}-\vec{u}_0) \alpha(\vec{u}_0),
\end{equation}
as a linear solution, where
\begin{equation}\label{eq12b}
\alpha(\vec{u})=\lim_{u\rightarrow 0} f(u, \vec{u}) u^{-1}, \quad  K_{\Delta_+=4}(u,\vec{u}-\vec{u}_0)=\frac{8}{\pi^2}
\frac{u^4}{\left[
u^2 +
(\vec{u}-\vec{u}_0)^2 \right]^4},
\end{equation}
and that
\begin{equation}\label{eq12c}
\left(\square_4 + m^2 \right)K_4(u,\vec{u}-\vec{u}_0)=0, \quad \left(-\square_u + m^2 \right) G(u,\vec{u};
w,\vec{w})=\delta(u,\vec{u};
w,\vec{w})/\sqrt{g_4},
\end{equation}
in which $K_4(u,\vec{u})$ and $G(u,\vec{u}; w,\vec{w})$ are the bulk-to-boundary and bulk-to-bulk propagators respectively,
and the
latter is given by
\begin{equation}\label{eq12d}
G(v)=\frac{1}{10 \pi^2} v^{-4} F\left(4, \frac{3}{2}; 6; -2v^{-1}\right), \quad v^{-1}=\frac{2 u
w}{(u-w)^2+(\vec{u}-\vec{w})^2},
\end{equation}
where $F(...)$ is the hypergeometric function, and one may note that $\frac{1}{1+v} \equiv \zeta$ is the "chordal distance"
between two
points in $EAdS$ space. Doing so, one obtains an approximate/perturbative solution that behaves near the boundary
($u\rightarrow 0$) as
\begin{equation}\label{eq13}
f(u,\vec{u}) \mapsto \tilde{\alpha}(\vec{u})\ u^{-1} + \tilde{\beta}(\vec{u})\ u^{+4};
\end{equation}
and we come back to this issue when discussing  the dual boundary solutions, with some proposals for
$\tilde{\alpha}(\vec{u})$
and
$\tilde{\beta}(\vec{u})$.

However, we note that the solution (\ref{eq10}) is indeed for the massless $\phi^4$ model and, because of the conformal
invariance, it is valid up to the conformal factor ($\Omega(u)$) for the tachyonic pseudoscalar in the bulk. Then, the
solution for the non-minimally coupled pseudoscalar in (\ref{eq11}) can be obtained by approximate methods, where some terms
(like the third one) of the equation role as a perturbation. Indeed, for the massive case, the mass term breaks the conformal
invariance softly, and one is not able to find a regular solution with finite action. The decay of the vacuum $f=0$ in the
case is dominated by the constrained instantons as the approximate solutions surveyed in \cite{Affleck1}, where a general
formalism for building them and evaluating the functional integrals is introduced, originally. In fact, with $m\neq 0$, the
massless solution is not exact, but for $b_0^2 m^2 \ll 1$ the constrained instanton behaves like the massless solution
(\ref{eq10}) in limit of $x_{\mu}\ll b_0$ and falls off exponentially for $x_{\mu}\gtrsim m^{-1}$ like that in a free massive
theory. In low energies, the small-size constrained instantons have the dominant contributions and the decay probability is
proportional to exponentials of the actions; look at \cite{KubyshinTinyakov01} for related studies. We also notice that these
constraints do not contradict with the condition $a_0> b_0 \geq 0$ to have the regular solution (\ref{eq10a}).

In addition, we note that there are some bulk instantons which break the $EAdS_4$ isometry $SO(4,1)$ down to $SO(3,1)$ or
$SO(4)$ and so an infinite family of the boundary instantons on $S^3$ with the same conformal symmetry, which may  in turn be
used to give a dual description of the cosmological singularities as in \cite{SmolkinTurok}, are accessible. We will find
some
instances of these boundary solutions for the non-minimally coupled case in subsection \ref{sub3.3}.

\subsection{The Action Corrections}
Now, we try to evaluate the corrections to the background action based on the solution for the conformally coupled case. The
appropriate part
of the 11d supergravity action with Euclidean signature in the case reads \footnote{One may also evaluate the action
(\ref{eq07}) on the solution (\ref{eq10a}) with a similar procedure and result.}
\begin{equation}\label{eq15}
  S_{11}^E = -\frac{1}{4 \kappa_{11}^2}  \int \left(\tilde{G}_4 \wedge \ast \tilde{G}_4 - \frac{i}{3} \tilde{\mathcal{A}}_3
  \wedge
  \tilde{G}_4 \wedge \tilde{G}_4 \right),
\end{equation}
where we use $\tilde{G}_4$ in (\ref{eq01b}) and that
\begin{equation}\label{eq15a}
\ast \tilde{G}_4 = \frac{8}{3} R^3 \bar{f}_1 J^3 \wedge e_7 - \frac{R^5}{2}  \ast_4 df \wedge J^2 + \frac{R^7}{8} f\
\mathcal{E}_4
\wedge J \wedge e_7,
\end{equation}
and
\begin{equation}\label{eq15b}
\tilde{G}_4=d\tilde{\mathcal{A}}_3, \quad \tilde{\mathcal{A}}_3 = \tilde{\mathcal{A}}_3^{(0)} + R^4 f\ J \wedge e_7, \quad
\tilde{G}_4^{(0)} = d\tilde{\mathcal{A}}_3^{(0)} = \bar{f}_1 \mathcal{E}_4,
\end{equation}
with $\bar{f}_1$ in (\ref{eq04}) and the conventions in (\ref{eq05a}). By plugging these into (\ref{eq15}), we have
\begin{equation}\label{eq15c}
 \begin{split}
  S_{11}^E = & -\frac{R^3}{4 \kappa_{11}^2} \int \left(\frac{8}{3} \bar{f}_1^2 \mathcal{E}_4 + \frac{R^6}{2} df \wedge \ast_4
  df +
  \frac{R^8}{8} f^2\  \mathcal{E}_4 \right) \wedge J^3 \wedge e_7 \\
  & + i\frac{R^8}{6 \kappa_{11}^2} \int \left(-f df \wedge \tilde{\mathcal{A}}_3^{(0)} + \bar{f}_1 f^2 \mathcal{E}_4 \right)
  \wedge J^3
  \wedge e_7,
 \end{split}
\end{equation}
with a note that by setting
\begin{equation}\label{eq15d}
\int f df \wedge \tilde{\mathcal{A}}_3^{(0)} = \frac{1}{2} \int \left( -f^2 \tilde{G}_4^{(0)} + d(f^2
\tilde{\mathcal{A}}_3^{(0)})
\right)
\end{equation}
in the last action, we get the equations (\ref{eq03a}) and (\ref{eq03b}) for $\bar{f}_1$ and $f$ respectively, and that the
second (surface) term above is a total derivative that do not affect the equations and is discarded throughout.

Now, with the conventions
\begin{equation}\label{eq15e}
    \bar{f_1}=\frac{3}{16} i R^5 f^2 - \frac{3}{8} i R^3, \quad  \mathcal{E}_4 = \frac{dx \wedge dy \wedge dz \wedge
    du}{u^4},
    \quad
    dVol(S^7/Z_k)= \frac{R^7}{3!} J^3 \wedge e_7,
\end{equation}
we arrive at
\begin{equation}\label{eq15f}
  S_{11}^E = S_0 - \frac{1}{4 \kappa_{11}^2} Vol(S^7/Z_k) \int \left(8 R^2  df \wedge \ast_4 df + \frac{7}{2} R^4 f^2
  \mathcal{E}_4 +
  \frac{9}{16} R^6 f^4 \mathcal{E}_4 \right),
\end{equation}
where $S_0=\frac{9}{\kappa_{11}^2 R^2} Vol(11d)$ is the contribution of the ABJM background. To advance, we use $df \wedge
\ast_4 df = d(f \ast_4 df) - f d(\ast_4 df)$ and the equation (\ref{eq03b}) to write the correction terms, in the unit 7d
internal volume, as
\begin{equation}\label{eq15g}
  S_{11}^{modi.} = - \frac{1}{4 \kappa_{11}^2} \int_{EAdS_4} \left(\frac{15}{2} R^4 f^2 \mathcal{E}_4 + \frac{39}{16} R^6 f^4
  \mathcal{E}_4 \right).
\end{equation}
Then, by plugging the solution (\ref{eq10a}) into this action and setting $\vec{u}_0=0$ for simplicity, which in turn
corresponds to
instantons with the size $b_0$ in origin of a boundary $S^3_\infty$, we finally obtain
\begin{equation}\label{eq16}
  S_{11}^{modi.} = - \frac{1}{24} \sqrt{\frac{3k^3}{R^5}} \left(60 b_0^2\ M1 -\frac{39}{9} b_0^4\ M2 \right),
\end{equation}
as the finite part of the correction, where $\kappa_{11}^2=\frac{16 \pi^5}{3} \sqrt{\frac{R^9}{3k^3}}$ is used \cite{ABJM}
and
\begin{equation}\label{eq16a}
M2=\frac{\left[5a_0^3 b_0^2 - 2a_0^5 - 3 a_0 b_0^4 + 2 (-b_0^2 + a_0^2)^{5/2}\right]}{b_0^4 (-b_0^2 + a_0^2)^{5/2}},
\end{equation}
\begin{equation}\label{eq16b}
M1=\frac{\left[a_0\ ln(2) - a_0\ ln\left(a_0 + (-b_0^2 + a_0^2)^{1/2} \right) + a_0\ ln\left(- b_0^2 + a_0^2\right) + (-b_0^2
+
a_0^2)^{1/2}\right]}{(-b_0^2 + a_0^2)^{3/2}},
\end{equation}
come from the second and the first term integration of (\ref{eq15g}), respectively. We note that the correction is small in
the legality
limit ($N\gg k^5$) of the model and finite $a_0$ and $b_0$. It is also notable that $\int f^2 \mathcal{E}_4 \rightarrow
\infty$
originally, whose singularity comes from the $u=0$ point. But, because of the renormalization arguments in \cite{Skenderis},
we have
considered the singularity at $\epsilon \geq 0$ instead of zero and then taken the finite part of the resulting action in the
limit of
$\epsilon\rightarrow 0$, while its infinite part is excluded as equal and opposite to the needed terms to neutralize it.

\section{The 3-Dimensional Field Theory Aspects}

\subsection{Basic Correspondence}
The primary fact is that for a (pseudo)scalar field, near the boundary ($u \rightarrow 0$) of the Euclidean $AdS_4$, we can
write the
asymptotic expansion \cite{KlebanovWitten}
\begin{equation}\label{eq18}
f(u,\vec{u})\thickapprox u^{\Delta_-} \alpha(\vec{u}) + u^{\Delta_+} \beta(\vec{u}),
\end{equation}
where $\Delta_+$ and $\Delta_-$, corresponding to normalizable and non-normalizable bulk modes respectively, are the smaller
and larger
roots of $m^2 L^2 = \Delta (\Delta-3)$, with $m$ for the field's mass. $\alpha$ and $\beta$ have the holographic
interpretations as the
vacuum expectation value (vev) and source for the boundary operator with the scaling dimension $\Delta_-$ and inversely for
the operator with $\Delta_+$. Such a (pseudo)scalar can be quantized with either Dirichlet($\delta \alpha=0$)- or Neumann
($\delta\beta=0$)- boundary condition, where the latter is used for the (pseudo)scalars with $-\frac{9}{4}<m^2 L^2<
-\frac{5}{4}$. These boundary conditions preserve
the asymptotic symmetry of $AdS_4$ and are consistent with two possible boundary conformal field theories.

Now, we note that the bulk objects here are pseudoscalars because they come from the internal ingredients of
$\mathcal{A}_{MNP}$. Then,
we look at the symmetries of the bulk ansatz (\ref{eq01b}) and a solution like (\ref{eq10a}). We note that, with considering
the internal
space as a $S^1/Z_k$ fibration on $CP^3$, the ansatz is a singlet of $SU(4) \times U(1)$ in that both $J$ and $e_7$ are
$SU(4)$ invariant
and do not carry any $U(1)$ charge and so, the corresponding boundary operators should have the same symmetry. Still, for the
c.c. case, the
conformal symmetry is preserved while for the n.m.c. case it is broken and so, the boundary solutions should
respects it as well. Further, from the ansatz, we see that the corresponding (anti)M2-branes wrap around the mixed internal
directions
and so they break all supersymmetries as it is confirmed with other arguments and a proof outlined in Appendix A.

On the other hand, one may remember that we have been considering the skew-whiffed background 4-form flux, which in turn
matches to the anti-M2-branes theory \cite{NilssonPope}, \cite{Duff}, \cite{Forcella.Zaffaroni}, to achieve the c.c.
pseudoscalar. In fact, one can have the desired singlet bulk modes when he/she exchanges the representations $\textbf{8}_s
\rightarrow \textbf{1}_{2} \oplus \textbf{1}_{-2} \oplus \textbf{6}_{0}$ and $\textbf{8}_c \rightarrow {\textbf{4}}_{-1}
\oplus \bar{\textbf{4}_{1}}$ of $SO(8)\rightarrow SU(4) \times U(1)$ for
the supercharges and fermions of the original theory respectively, while $\textbf{8}_v \rightarrow \textbf{4}_{1} \oplus
\bar{\textbf{4}}_{-1}$ is for scalars. So we propose $\textbf{8}_c \leftrightarrow \textbf{8}_s$, and then the pseudoscalars
set in
$\textbf{35}_s \rightarrow \textbf{1}_{0} \oplus \bar{\textbf{1}}_{4} \oplus \textbf{1}_{-4} \oplus \bar{\textbf{6}}_{2}
\oplus
\textbf{6}_{-2}\oplus \bar{\textbf{20}}_{0}$, the scalars in $\textbf{35}_{v} \rightarrow \textbf{10}_{2} \oplus
\bar{\textbf{10}}_{-2}
\oplus \textbf{15}_{0}$ and gauge bosons in $\textbf{28} \rightarrow \textbf{1}_{0} \oplus \bar{\textbf{6}}_{2} \oplus
\textbf{6}_{-2}
\oplus \textbf{15}_{0}$ remain unchanged. Therefore, the state may be interpreted as adding some M2-branes to the
skew-whiffed
background
(anti-M2-branes) theory as the resulting theory is for anti-M2-branes, with breaking all supersymmetries. For the n.m.c.
pseudoscalar,
for which we have just Wick rotated the original 4-form flux (\ref{eq002}), the argument is similar. Indeed, one may propose
some
anti-M2-brane to be added on top of the original M2-branes as the resultant state has the same symmetries as the former and
so, the
resulting theory is also for anti-M2-branes with the same swapping of representations. \footnote{It is notable that besides
$m^2
L^2=+4$, we could have another massive $m^2 L^2=+18$ mode if we set $\bar{C}=\frac{17}{8}$ from (\ref{eq06}). The recent
couple is also
found in \cite{Gauntlett03} upon some consistent Kaluza-Klein reduction in a similar context, and are interpreted as
\emph{squashing} and
\emph{breathing} (pseudo)scalar in the lower dimension; look also in \cite{LiuSati} for some related discussions.} In the
following
subsections we concentrate on the field theory counterparts for both cases.

\subsection{The Conformally Coupled Case: Dual Instantons} \label{sub3.2}
We note that for the pseudoscalar $m^2 L^2=-2$, the conformal dimensions are $\Delta_{\mp}=1,2$. We have already used the
quantization
with Dirichlet boundary condition in the case \cite{I.N} and found a dual solution and an agreeing $\Delta_+=2$ operator.
Here
we do the
same job for the operator $\Delta_-=1$ corresponding to quantization with Neumann or mixed boundary condition. On the other
hand, we note
that the scalar theories coupled to gravity, with the scalar masses around the so-called Breitenlohner-Freedman bound
\cite{BreitenlohnerFreedman} $m_{BF}^2 L^2\geq -\frac{9}{4}$ in $AdS_4$, admit a large class of boundary conditions and are
always called
the \emph{designer gravity} theories in that their properties are depended on the choices of the boundary conditions. The
deformation
here is a \emph{triple-trace deformation} that destabilizes the classical gravity solution with the false vacuum decay,
resulting
in a \emph{big-crunch} in $AdS_4$ as well; Look at \cite{HertogMaeda01}.

Therefore, starting with Neumann boundary condition, we change it with a one-parameter deformation as
\begin{equation}\label{eq19a0}
 \beta = - L \hat{h} \alpha^2,
\end{equation}
where the $L$ factor is for convenience and $\hat{h}$ as the deformation parameter labels various boundary conditions. So,
with the
\emph{mixed} boundary condition, we should correct the boundary action with the term
\begin{equation}\label{eq19}
-S_{on} = \int d^3\vec{u}\ \mathcal{O}_2(\vec{u}) \mathcal{O}_1(\vec{u}) = - \frac{L \hat{h}}{3} \int d^3\vec{u}\
\mathcal{O}_1(\vec{u})^3 = W,
\end{equation}
where $S_{on}$ and $W$ are the corresponding bulk on-shell action and the boundary generating-functional respectively, and
\begin{equation}\label{eq19a}
\frac{1}{3} \beta(\vec{u})= - \frac{\delta W[\alpha(\vec{u})]}{\delta\alpha(\vec{u})}= \langle \mathcal{{O}}_2
\rangle_{\alpha} \equiv
\sigma(\vec{u}), \quad \alpha(\vec{u}) = - \frac{\delta W[\sigma(\vec{u})]}{\delta\sigma(\vec{u})}= - \langle \mathcal{{O}}_1
\rangle_{\beta},
\end{equation}
with $\mathcal{O}_1$ and $\mathcal{O}_2$ for the dimension-1 and -2 boundary operators, respectively. \\
On the other hand, from the Taylor’s expansion of the solution (\ref{eq10a}) around $u=0$, we have
\begin{equation}\label{eq20}
\alpha(\vec{u}) = \frac{4}{R \sqrt{\lambda}} \frac{b_0}{\left[-b_0^2 + a_0^2 + (\vec{u}-\vec{u}_0)^2\right]}, \quad
\beta(\vec{u}) =
-\frac{8}{R \sqrt{\lambda}} \frac{a_0 b_0}{\left[-b_0^2 + a_0^2 + (\vec{u}-\vec{u}_0)^2\right]^2},
\end{equation}
and so $\hat{h}= \sqrt{\lambda}\ \frac{a_0}{b_0}$ from (\ref{eq19a0}), with $\lambda=3$ for both cases here. One may also
evaluate the integral
in (\ref{eq19}) whose finite contribution reads
\begin{equation}\label{eq19bb}
S_{on}^{modi.} = - \frac{8}{3} \frac{\pi^2}{\lambda R^2} \frac{a_0 b_0^2}{\left(a_0^2 - b_0^2 \right)^{3/2}}, \quad
\int_0^\infty
\frac{r^2}{\left(- b_0^2 + a_0^2 + r^2\right)^3} dr = \frac{\pi}{16 \left(a_0^2 - b_0^2\right)^{3/2}}.
\end{equation}

Next, what is the plain form for the boundary dimension-1 operator? According to above symmetry arguments, the operators must
be $SU(4)_R
\times U(1)_b$-singlet as we have the same singlet bulk pseudoscalar. Following the arguments in \cite{I.N}, as we know there
is not any
singlet dimension-1 operator in the original ABJM model and so, as an alternative, because of the skew-whiffing $\textbf{8}_s
\leftrightarrow \textbf{8}_c$, we use the singlet fermion $\psi$ now in $\textbf{8}_s \rightarrow \textbf{1}_{2} \oplus
\textbf{1}_{-2} \oplus \textbf{6}_{0} \rightarrow \textbf{8}_c$ to make the wished operator as $\mathcal{{O}}_1 =
\left(tr(\psi
\bar{\psi})\right)^{1/2}$. Then, by setting the scalars to zero, deforming the remaining part of the ABJM boundary action
with
$W$ in
(\ref{eq19}) and following the procedure in \cite{Me5}, we simply obtain the solution
\begin{equation}\label{eq21}
      \psi = \sqrt{\frac{N}{\lambda}} \frac{12 b_0 b_1}{R a_0} \frac{\left(b_1 + i (x - x_0)^i \gamma_i \right)}{\left(b_1^2
      +
      (x -
      x_0)_i (x - x_0)^i \right)^{3/2}} \left(\begin{array}{c}   1  \\   0   \end{array}\right),
\end{equation}
where $b_1, b_2,..$ are some boundary constants, and we have used the ansatz $\psi_{\hat{a}}^a= \frac{\delta_{\hat{a}}^a}{N}
\psi$ equivalent to
concentrating on $U(1) \times U(1)$ part of the gauge group with $A_i^\pm \equiv (A_i \pm \hat{A}_i)$; and then setting
$A_i^-
= 0$ left
us with a self-interacting spinor corresponding to the self-interacting pseudoscalar in the bulk.

Then, we notice that
\begin{equation}\label{eq22}
     \langle \mathcal{O}_{1} \rangle_{\beta} \sim \left(b_1^2+(\vec{u}-\vec{u}_0)^2 \right)^{-1} \sim \alpha(\vec{u}),
\end{equation}
with $b_1^2 = -b_0^2 + a_0^2 $, confirming the bulk/boundary correspondence. In addition, one may note that the vev of the
operator
diverges in the large $N$ limit, which is in turn typical of the field theory dual to describe a big crunch in the bulk.
Meanwhile, the
finite part of the boundary action based one the solution (\ref{eq21}) becomes
\begin{equation}\label{eq23}
   S_{corr.} = \frac{R \hat{h}}{12} \int d^3\vec{u}\ \left(tr(\psi \bar{\psi})\right)^{3/2} = \sqrt{\frac{N^3}{\lambda^2}}
   \left(\frac{6
   \pi b_0}{R a_0} \right)^2, 
\end{equation}
where we have considered the instantons in the center ($\vec{u}_0=0$) of $S_\infty^3$ with radius $r$ and the same
integration
formula in
(\ref{eq19bb}).

In addition, we note that for the mixed boundary condition, the dual operator is $\Delta_- = +1$; and that any solution to
the
bulk AdS
equation should be dual to an extremum or a vacuum of the \emph{effective action} of the dual boundary CFT. On the other
hand,
we are
aware of the following dictionary
\begin{equation}\label{eq24}
   \Gamma_{eff.} [\sigma] = \tilde{\Gamma}_{eff.} [- \alpha] - \int d^3\vec{u}\ \sigma(\vec{u}) \alpha(\vec{u}), \quad
   \Gamma_{eff.}
   [\sigma] = W[\sigma], \quad \tilde{\Gamma}_{eff.} [- \alpha] = - W[\alpha],
\end{equation}
where $\Gamma_{eff.} [\sigma]$ and $\tilde{\Gamma}_{eff.} [\alpha]$ are the effective actions of the usual (with $\Delta_+ =
2$) and dual
(with $\Delta_- = 1$) CFT respectively, which are indeed connected by a Legendre transform. Evaluating the boundary effective
action in
the case needs a special effort as done in \cite{Papadimitriou02}, for the dual CFT deformed by (\ref{eq19a0}), as
\begin{equation}\label{eq25}
 \begin{split}
   \tilde{\Gamma}_{eff.} [\alpha] = \int d^3\vec{u}\ & \left(V_{eff.}(\alpha) + \frac{1}{12 \sqrt{\lambda}}
   \left(\frac{6}{L^2} \alpha +
   2 \alpha^{-1} \partial_i \alpha \partial^i \alpha \right) \right), \\
  & \ \ \ \ V_{eff.}(\alpha) =\frac{1}{3} \left(\sqrt{\lambda} - \hat{h} \right) \alpha^3,
 \end{split}
\end{equation}
in a two-derivative approximation, where $V_{eff.}(\alpha)$ is the holographic effective potential. We note that the
instanton
solution (\ref{eq10a}) is an extremum of the all-order effective action although a similar two-derivative boundary action
like
(\ref{eq25}) gives that solution.

Further, we have noticed that the bulk instanton solution is regular when $a_0 > b_0 \geq 0$ and so $\hat{h} > \sqrt{\lambda}
> 0$ that
results in $V_{eff.}(\alpha) <0$, which in turn means the effective potential is unbounded from below. The latter bodes that
the
instantons mediate the quantum tunneling of the conformal vacuum (the local minimum of $V_{eff.}(\alpha)$) at $\alpha = 0$
because of
the instability imposed by the (marginal) triple-trace deformation (\ref{eq19a0}). Then, one may estimate the \emph{rate of
decay} or
decay probability for the conformal-vacuum as
\begin{equation}\label{eq26}
  \mathcal{P} \sim e^{- \tilde{\Gamma}_{eff.}}|_{inst.} \sim e^{S_{inst.}},
\end{equation}
where $S_{inst.} = S_{11}^{modi.} + S_{on}^{modi.}$. In fact, one should note that because we do not know the exact form of
the effective
action for the boundary theory, to evaluate $\mathcal{P}$, we have alternatively used the correction of the bulk action based
on the exact
solution (that is $S_{11}^{modi.}$ in (\ref{eq16})) plus the boundary contribution (that is $S_{on}^{midi.}$ in
(\ref{eq19bb})); and so
we have a probability for the decay.

\subsection{The Non-Minimally Coupled Case: Boundary Solutions}  \label{sub3.3}
For the pseudoscalar $m^2 L^2=+4$ from the equation (\ref{eq06}), with corresponding boundary operators $\Delta_{\mp}= -1,
+4$, we look at the
normalizable bulk mode along with Dirichlet boundary condition ($\delta \alpha=0$) with respect to (\ref{eq18}). The dual
$SU(4)_R
\times U(1)_b$-singlet dimension-4 operator can be formed according to the known 11d supergravity spectrum over $AdS_4 \times
S^7/Z_k$
\cite{NilssonPope}, \cite{ABJM} and the appropriate skew-whiffing $\textbf{8}_s \leftrightarrow \textbf{8}_c$. In fact, for
the $0^{-(1)}$
pseudoscalars, the proposed operator in the case can be \cite{EDHoker02}, \cite{Bianchi2}, \cite{Forcella.Zaffaroni}
\begin{equation}\label{eq27}
     \acute{\mathcal{O}}_4 = tr \left(\Psi^{[I} \Psi^{J^\dagger]} X^{[K} X_{[L}^\dagger X^{L]} X_{K]}^\dagger \right),
\end{equation}
where $X^I \rightarrow (Y^A, Y_A^\dagger)$ with $A=1,2,3,4$ and $\Psi^I \rightarrow (\psi^n, \psi, \bar{\psi})$ with
$I,J...=(1,...6,7,8)=(n,7,8)$, $\psi=\psi^7+i\psi^8$, $\psi^\dagger=\bar{\psi}$ transform in the representation
$(\textbf{4}_1,
\bar{\textbf{4}}_{-1})$ and $(\textbf{6}_{0}, \textbf{1}_{2}, \textbf{1}_{-2})$ of $SO(8)\rightarrow SU(4)_R \times U(1)_b$,
after the
skew-whiffing, respectively. One should also note to the suitable trace subtractions in the symmetrized products of
(\ref{eq27}).
Besides, it is noticeable that the operator $\acute{\mathcal{O}}_4$ might be made of the dimension-3 operator from $X^I$'s,
which is in turn proportional with the ABJM and BLG scalar potentials \cite{Raamsdonk}. Indeed, the second generation
(descendants) of the
$\acute{\mathcal{O}}_3$ operator \cite{Me4} gives the suitable operator. By the way, we employ the plain form
\begin{equation}\label{eq27a}
     \mathcal{O}_4 = tr \left(\psi_A \psi^{A\dagger} Y^B Y_B^\dagger Y^C Y_C^\dagger \right).
\end{equation}
Then, we note that with just the singlet ($\textbf{1}_{2}$) fermion $\psi$ and the scalars in the original representation, we
have the
suitable singlet operator in $\textbf{1}_{2} \otimes \textbf{1}_{-2} \otimes \textbf{4}_1 \otimes \bar{\textbf{4}}_{-1}
\otimes
\textbf{4}_1 \otimes \bar{\textbf{4}}_{-1}$.

On the other hand, similar to that in (\ref{eq19a}), we can write
\begin{equation}\label{eq28}
\frac{1}{5} \langle \mathcal{{O}}_4 \rangle_{\tilde{\alpha}} = - \frac{\delta
\tilde{W}[\tilde{\alpha}(\vec{u})]}{\delta\tilde{\alpha}(\vec{u})} = \tilde{\beta}(\vec{u}) \Rightarrow \tilde{W} = -
\frac{1}{5} \int
d^3\vec{u}\  \tilde{\alpha}(\vec{u}) \mathcal{O}_4(\vec{u}),
\end{equation}
as the boundary deformation term. Now, next to the singlet fermion, we use just one scalar with the ansatzs
\begin{equation}\label{eq29}
      \tilde{\psi}_{\hat{a}}^a= \frac{\delta_{\hat{a}}^a}{N} \tilde{\psi}, \qquad Y = \frac{\tilde{h}(r)}{N} I_{N\times N},
\end{equation}
where $\tilde{h}(r)$ is for a scalar profile on the boundary and $I_{N\times N}$ is the unit matrix. Therefore, in the
boundary action
(see \cite{Me3}, \cite{ Me4}), the fermion and boson potentials vanish and the deformed Lagrangian reads
\begin{equation}\label{eq30}
  \mathcal{L}_{def.} = \mathcal{L}_{CS} + \hat{\mathcal{L}}_{CS} - tr \left(D_k Y^{\dagger} D^k Y \right) - tr
  \left(\psi^{\dagger} i
  \gamma^k D_k \psi \right)- \frac{1}{5} \tilde{\alpha}(\vec{u})\ tr \left((\bar{\psi} \psi) (Y^{\dagger} Y)^2 \right),
\end{equation}
where $D_k \Phi =\partial_k \Phi + i A_k \Phi - i \Phi \hat{A}_k$, with $\Phi$ for both $Y$ and $\psi$, and
\begin{equation}\label{eq30a}
      \mathcal{L}_{CS} = \frac{i k}{4\pi}\ \varepsilon^{k ij} \ tr \left(A_i \partial_j A_k + \frac{2i}{3} A_i A_j A_k
      \right), \quad
      F_{ij} = \partial_i A_j - \partial_j A_i + i [A_i, A_j],
\end{equation}
and the same expression for $\hat{\mathcal{L}}_{CS}$ by changing $A$ to $\hat{A}$. Then, from the action, the equations for
the scalar
and fermion read
\begin{equation}\label{eq31a}
      D_k D^k Y - \frac{2}{5} \tilde{\alpha}(\vec{u})\  tr(\bar{\psi} \psi)\ tr(Y^{\dagger} Y) Y = 0,
\end{equation}
\begin{equation}\label{eq31b}
     i \gamma^k D_k \psi + \frac{1}{5} \tilde{\alpha}(\vec{u})\ tr(Y^\dagger Y)^2 \psi=0,
\end{equation}
respectively and because of the ansatz (\ref{eq29}), we are indeed working with $U(1) \times U(1)$ part of the complete gauge
group and with $A_i^\pm
\equiv (A_i \pm \hat{A}_i)$, the equations for $A_i$ and $\hat{A}_i$ become those written in \cite{Me3} and \cite{Me5}, where
$F_{ij}^- =
0 \Rightarrow A_i^- = 0$ left us with a self-interacting fermion field. From the coupled equations (\ref{eq31a}) and
(\ref{eq31b}), we
simply get
\begin{equation}\label{eq32a}
     \partial_k \partial^k \tilde{h}(r) = 0 \Rightarrow \tilde{h}(\vec{u}, \vec{u}_0) = \tilde{b}_1 + \frac{\tilde{b}_2}{\mid
     \vec{u} -
     \vec{u}_0 \mid},
\end{equation}
\begin{equation}\label{eq32b}
     i \gamma^k \partial_k \psi = 0 \Rightarrow \psi= \frac{\sqrt{N}}{2} i \sqrt[3]{\frac{4}{5}} \frac{(\vec{u} - \vec{u}_0).
     \vec{\gamma}}{\big[(\vec{u} - \vec{u}_0)^2 \big]^{3/2}} \left(\begin{array}{c}
                                                                                                                  1 \\
                                                                                                                  0
                                                                                                                \end{array}\right),
\end{equation}
where $\vec{\gamma} \equiv (\sigma_2, \sigma_1, \sigma_3)$. \footnote{We remember the condition (\ref{eq02d}) to satisfy the
bulk
equations, where $f_2$ could be for a massless scalar. It is notable that the solution $\tilde{h}(r)$ can be matched with
that
one. Indeed,
if we set also the fermions to zero and keep just the boundary scalar in (\ref{eq29}), we can construct the dual dimension-3
operator in the
original representation of the ABJM model, as done in \cite{Me4} and \cite{Me5}.} From these solutions, we note that vev of
the
operator reads
\begin{equation}\label{eq33}
     \langle \mathcal{O}_{4} \rangle_{\tilde{\alpha}} \sim {\big(\tilde{b}_3^2 + (\vec{u}-\vec{u}_0)^2\big)^{-4}} \sim
     \tilde{\beta}(\vec{u}),
\end{equation}
with $\tilde{b}_3 =0$ here, which matches with the boundary behavior of the bulk pseudoscalar from (\ref{eq13}) and is
consistent
with (\ref{eq12b}) with the constraints hinted at the end of subsection \ref{sub2.2}. Next, with the equations, the
correction
to the
action from (\ref{eq30}) based on the solutions, becomes
\begin{equation}\label{eq34}
   \tilde{S}_{modi.}  = - \int_{S^2} d^2\vec{u} \ \tilde{h} (\partial_k \tilde{h}) \Rightarrow \tilde{S}_{modi.}^{inst.} =
   4\pi
   \tilde{b}_1 \tilde{b}_2,
\end{equation}
with the boundary as a 3-sphere at infinity concentrated around $\vec{u}_0=0$ and a similar procedure in \cite{N}, where the
finite
contribution is from infinity.

Another alternative boundary solution is accessible with $\mathcal{O}_{2}=tr(\psi \bar{\psi})$ and a \emph{double-trace}
deformation as
$\mathcal{O}_{4}=\mathcal{O}_{2}^2$. Then, with the same skew-whiffing and setting the scalars to zero, the resultant
equation
for $\psi$ reads
\begin{equation}\label{eq35}
     i \gamma^k D_k \psi + \frac{2}{5} \tilde{\alpha}(\vec{u})\ tr(\psi \bar{\psi}) \psi = 0.
\end{equation}
To find a solution, we use a similar ansatz as in the previous subsection and \cite{I.N} and also \cite{Me5} with the
solution
\begin{equation}\label{eq36}
      \psi = \pm \sqrt{\frac{15 \tilde{b}_3 N^2}{2}} \frac{\left(\tilde{b}_3 + i (x-x_0)^k \gamma_k
      \right)^\varsigma}{\left(\tilde{b}_3^2 + (x-x_0)_k (x-x_0)^k \right)^2 \left(\tilde{b}_3^{\dagger} - i (x-x_0)_k
      \gamma^{k \dagger}
      \right)^{\varsigma/2}} \left(\begin{array}{c}   1  \\     0   \end{array}\right),
\end{equation}
from which, with $\varsigma = 4$, we confirm the correspondence mentioned in (\ref{eq33}). Meanwhile, the action value based
on this
solution reads
\begin{equation}\label{eq37}
   \tilde{S}_{modi2.} = \frac{\alpha}{5} \int d^3\vec{u}\ \big(tr(\psi \bar{\psi})\big)^{2}, \quad \int_0^\infty
   \frac{d^3\vec{u}}{\big(\tilde{b}_3^2+(\vec{u}-\vec{u}_0)^2\big)^4} = \frac{\pi}{32 \tilde{b}_3^5}, \quad
    \tilde{S}_{modi2.}^{inst.} = \frac{225}{160} \frac{\pi^2 N^4}{\tilde{b}_3^3},
\end{equation}
where $\tilde{b}_3 > 0$ and the integration is done on $S_\infty ^3$ around $\vec{u}_0=0$ as usual. It is remarkable that the
boundary
solutions here may be well matched with the bulk when the constrained approximations outlined in subsection \ref{sub2.2} are
employed.
Anyway, the solutions in the case are some proposals to be identified with the bulk solutions of the equation (\ref{eq11}).
Meanwhile, one may be tempted to construct the $\Delta_- = -1$ operator similarly.

\section{Further Discussions}
In this paper, we have focused on two non-minimally coupled pseudoscalars in the bulk of Euclidean $AdS_4$ among a tower of
massive and tachyonic
(and also massless) modes from a general 4-from ansatz with keeping the prime geometry unchanged. Fortunately, the modes are
in the
spectrum of 11d supergravity over $AdS_4 \times S^7/Z_k$ when the internal space is considered as a $S^1/Z_k$ fiber bundle on
$CP^3$.
Indeed, because both pseudoscalars are $SU(4) \times U(1)$ singlet and break all 32 original supersymmetries, we should
exchange,
respectively, the representations $\textbf{8}_s$ and $\textbf{8}_c$ of the supercharges and fermions of the main M2-branes
theory
\cite{ABJM} to have the singlet $\textbf{1}_0$ now in $\textbf{35}_s \rightarrow \textbf{1}_{0} \oplus \bar{\textbf{1}}_{4}
\oplus
\textbf{1}_{-4} \oplus \bar{\textbf{6}}_{2} \oplus \textbf{6}_{-2}\oplus \bar{\textbf{20}}_{0}$.  Therefore, we proposed that
the resultant
boundary theory was for anti-M2-branes obtained with the same skew-whiffing of the original $\mathcal{N}=6$ conformal
Chern-Simon-matter 3d
field theory.

Then, with conformal flatness of the external space, we represented an exact instanton solution for the conformally coupled
case
($\xi=1/6$) with calculating its correction to the action, and also proposed a dual boundary solution based on a so-called
triple-trace
deformation \cite{Papadimitriou02}. In addition, as it is known that the instanton solution mediates some tunneling
processes,
an
estimate of the decay rate was also provided. But, for the non-minimally coupled massive pseudoscalar ($\xi=-1/3$), the
conformal invariance was
softly broken and one could not found an exact solution and so, the approximate solutions such as constrained instantons
\cite{Affleck1}
could be adjustable. Then, we proposed a dual $\Delta_+ =4$ operator and deformed the action with a suitable Dirichlet
boundary condition/term to get a plain boundary solution.

Although \emph{supersymmetry} breaking by the solutions is obvious in that the included (anti)M-branes, which source the
matching 4- and
7-from fluxes, wrap around some mixed internal and external directions; meanwhile the checking of supersymmetry for the
ansatz
(\ref{eq01b}) can be done, for instance, based on the integrability condition, like the procedure done in \cite{NilssonPope},
as we have
hinted it in appendix A and left the details for future studies. The \emph{instability} of the conformally coupled solution
is
described in \cite{deHaro2}, \footnote{It should be noted here that one may consider the Einstein equations (\ref{eq09}) with
the effective gravitational constant $\kappa_{eff.}^2 = (1- \xi \kappa_4^2 f^2)^{-1} \kappa_4^2$ and then, to have the
attractive gravitational force, one must impose $\left(1- \xi \kappa_4^2 f^2 \right) \geq 0$ as the extrema of the potential,
for the background with the negative cosmological constant, become $f = \pm \sqrt{\frac{3}{4 \pi \mathcal{G}_4}}$. On the
other hand, the stable $AdS_4$ vacuum $f =0$ becomes unstable under a small perturbation, caused by the mixed boundary
condition that we have considered in subsection \ref{sub3.2} for the conformally coupled case. Indeed, the unboundedness of
the boundary effective potential from below is interpreted as instability of dual theory against the marginal deformation
(\ref{eq19}) and as a result the pseudoscalar can tunnel from the local minimum at $\alpha=0$ into the instability region at
$\alpha\rightarrow \infty$ (see the figure 1 of \cite{Papadimitriou02})-- For an original study on instability in similar
situations look at \cite{ColemanDeLuccia} and at \cite{Shin'ichi Nojiri} for a quantum instability analysis of AdS
backgrounds.} while for multi-trace deformations and minimally and also non-minimally coupled (pseudo)scalars, one may do
similar analysis as, for instance, those in \cite{LucaVecchi} suitable here as well. For the massive $m^2 = +4$ pseudoscalar,
searching for a suitable \emph{bulk instanton} solution, about constrained instantons, symmetry groups or another approximate
ways such as \emph{valley instanton} method in \cite{hep-th/9512064}, will be interesting. The issue of \emph{backreaction}
of
the solutions on the background geometry is remarkable and needs to be investigated further--For a holographic
renormalization
of the irrelevant deformations, look at \cite{BaltvanRees}. It is also interesting to match the spectra here with the bulk
supergravity modes in \cite{Gauntlett03} and study their applications in other physical phenomena, such as cosmology and
superconductivity.

As a final point, it should be noted that we are indeed not aware of any non-renormalization theorem that guarantees the matching of the dimensions computed here in weak/strong limits of gravity/gauge theories. Therefore, we have actually used the free field theory assumption in proposals for the operators. In other words, it is known that the scaling dimension of chiral primary (or “short”) operators and their descendants are non-renormalized or protected against quantum corrections. But, with the supersymmetry breaking non-BPS operators, employed here as well, the latter statement is not valid in general as the operators may obtain anomalous dimensions after renormalization. In fact, by including the operators that break supersymmetry next to having nonperturbative effects, the non-renormalizablilty need to be explored further. For more information on the issue, see \cite{Seiberg} as an original related study and look, for instance, at \cite{hep-th/0009106}, \cite{hep-th/0010137} for non-renormalization theorems and corrections for operators and multi-point correlation functions of gauge-invariant composite operators in AdS$_5$/CFT$_4$ correspondence. In addition, for discussions on renormalizablilty of Chern-Simon-matter theories, look at \cite{Avdeev1993}, \cite{hep-th/9401053} and see \cite{0806.3951} for a discussion on anomalous dimensions of some operators in the ABJM model.

\begin{appendices}
\section{A Hint On Supersymmetry}
The main supersymmetry is broken, with the solutions, for some reasons. The first reason is the skew-whiffing that except for
$S^7$
breaks all supersymmetries \cite{Duff}. The second reason is the structure of the ansatz where the associated (anti)M-branes
wrap around
some mixed directions in the internal and external spaces and thus break supersymmetry completely. The third reason arises
from the multi-trace
deformation applied, at least for the conformally coupled case, as it is known that it breaks supersymmetries as well
\cite{HertogMaeda01}. Here, we hint on the direct methods for supersymmetry checking given the ansatz (\ref{eq01b}).

In fact, having a classical solution, the \emph{killing spinors} $\epsilon$ control the numbers of supersymmetries, which are
in turn
given by the solutions of the equations
\begin{equation}\label{eqa01}
     \delta\Psi_M \equiv \tilde{D}_M \epsilon = D_M \epsilon -\frac{1}{128} \left(\Gamma_M^{PQRS} - 8 \delta_M^P \Gamma^{QRS}
     \right)
     G_{PQRS} \epsilon,
\end{equation}
in which $\Psi_M$ is for the gravitino as a 32-component Majorana spinor, and
\begin{equation}\label{eqa02}
 \begin{split}
   & \ \ \ \ \ \ D_M \epsilon = \partial_M \epsilon + \frac{1}{4} \omega_M^{AB} \Gamma_{AB} \epsilon, \quad \left[D_M, D_N
   \right] =
   -\frac{1}{4} \mathcal{R}_{MN}^{AB} \Gamma_{AB} \epsilon, \\
   & G_4 = \frac{1}{4} G_{MNPQ} dX^{MNPQ}, \quad dX^{MNPQ} \equiv dX^M \wedge dX^N \wedge dX^P \wedge dX^Q,
 \end{split}
\end{equation}
where $\mathcal{R}_{MNPQ}$ is for the Riemann curvature tensor, $A,B,..$ here are for the 11d tangent (flat) space indices,
and the
high-dimensional gamma matrices as $\Gamma^{M_1 M_2 ... M_n}$ includes $n!$ anti-symmetrized terms with an overall $1/n!$
factor in the
front. Also, the spin-connection 1-forms $\omega_M^{AB}$ are the solutions to the equation
\begin{equation}\label{eqa03}
   T^A = de^A + \omega_B^A \wedge de^A, \quad e^A = e_M^A dX^M, \quad \omega_B^A = \omega_{MB}^A dX^M,
\end{equation}
where $T_{MN}^A$ is a vector-valued 2-form and that, from the metric compatibility and torsion freeness, one can obtain the
spin
connections for the metric (\ref{eq001}), with the suitable selected vielbeins such as those in \cite{Me5},
straightforwardly.

To continue, one should write the ansatz (\ref{eq01b}) in components that is
\begin{equation}\label{eqa04}
       G_{MNPQ} \equiv G_{{\mu}{\nu}{\rho}{\sigma}} + G_{{\mu}{11}{m}{n}} + G_{{m}{n}{p}{q}} = \bar{f}
       \mathcal{E}_{{\mu}{\nu}{\rho}{\sigma}} + \partial_{{\mu}} f J_{m n} + f \Omega_{mnpq},
\end{equation}
where $\Omega_{mnpq} \equiv 2 (J_{m n}J_{p q} - J_{p n}J_{m q} - J_{q n}J_{p m})$. Also, one may decompose the 11d gamma
matrices into
the external and internal components (look at \cite{LuPopeRahmfeld}) with some special conventions. Then, by solving the
killing spinor
equations (\ref{eqa01}), one can get the numbers of unbroken supersymmetries. Indeed, if we write the eleven equations from
vanishing the
supersymmetry variation $\delta\Psi_M =0$, with respect to the equation of motion (\ref{eq06}), see that all projections
imposed on $\epsilon$ vanish,
which in turn means that no spinor is preserved or that all supersymmetries are broken. Still, one can get the maximum
numbers
of
preserved supersymmetries from the \emph{integrability condition} similar to the method carried out in \cite{NilssonPope}.
The
details of
the supersymmetry checking and similar procedure to the latter, in the case, need more time and space and therefore we leave
them to
future studies.
\end{appendices}

\end{document}